\documentclass[eqsecnum,showpacs,preprint,amsmath,aps,nofootinbib]{revtex4-1}
\usepackage{amssymb}
\usepackage{gensymb}
\usepackage{amsmath}

\begin{document}

\title{Quantum Effects on all Lagrangian Points and Prospects to Measure Them in the Earth-Moon System}
\author{Emmanuele Battista}

\address{Dipartimento di Fisica, Complesso Universitario di Monte S. Angelo, \\
Istituto Nazionale di Fisica Nucleare, Sez. Napoli, Complesso Universitario di Monte S. Angelo,\\
Via Cintia Edificio 6, 80126 Napoli, Italy\\
 E-mail: ebattista@na.infn.it}

\author{Giampiero Esposito} 

\address{Istituto Nazionale di Fisica Nucleare, Sez. Napoli, Complesso Universitario di Monte S. Angelo,\\
 Via Cintia Edificio 6, 80126 Napoli, Italy \\
 E-mail: gesposit@na.infn.it}

\author{Simone Dell'Agnello}

\address{Istituto Nazionale di Fisica Nucleare, \\
Laboratori Nazionali di Frascati, 00044 Frascati, Italy \\
E-mail: simone.dellagnello@lnf.infn.it}

\author{Jules Simo}

\address{Department of Mechanical and Aerospace Engineering, University of Strathclyde, \\
Glasgow, G1 1XJ, United Kingdom\\
E-mail: jules.simo@strath.ac.uk}

\begin{abstract}
The one-loop long distance quantum corrections to the Newtonian
potential imply tiny but observable effects in the restricted three-body problem of celestial mechanics, i.e.,
both at the Lagrangian points of stable equilibrium and at those of unstable equilibrium the Newtonian values of planetoid's coordinates are changed by a few millimetres in the Earth-Moon system. First, we find that the equations governing the position of both noncollinear and collinear quantum libration points are algebraic fifth degree and ninth degree equations, respectively. Second, we discuss the prospects to measure, with the help of laser ranging, the above departure from the equilateral triangle picture, which is a challenging task. On the other hand, a modern version of the planetoid is the solar sail, and much progress has been made, in recent years, on the displaced periodic orbits of solar sails at all libration points. By taking into account the quantum corrections to the Newtonian potential, displaced periodic orbits of the solar sail at libration points are again found to exist.
\end{abstract}

\keywords{Restricted three-body problem, Effective field theories, One-loop quantum corrections, Laser ranging, Displaced periodic orbits.}

\maketitle

\section{Introduction}

One of the most outstanding problems of modern theoretical physics is represented by the incompatibility of quantum mechanics and general relativity, which results in a perturbatively non-renormalizable theory of quantum gravity. Despite that, quantum predictions can be made in non-renormalizable theories by employing the techniques of effective field theory, which allow a natural separation of the effects due to low-energy physics from those of the (unknown) high energy theory. General relativity fits naturally into the context of effective field theory as gravitational interactions are proportional to energy and are easily organized into an energy expansion, where the expansion scale factor is the Planck length $l_{\mathrm{P}}$. The leading (i.e., one-loop) long distance quantum corrections to the Newtonian potential are entirely ruled by the Einstein-Hilbert part of the full action functional and, besides being parameter free (apart from the Newton constant $G$), dominate over other quantum predictions in the low-energy limit \cite{D94,D03}. These corrections are the first modification due to quantum mechanics and are organized in inverse powers of the distance. As they are independent of the eventual high energy theory of gravity, depending only on the massless degrees of freedom and their low-energy couplings, they are true predictions of quantum gravity. More precisely, the Newtonian potential between two bodies of masses
$m_{\mathrm{A}}$ and $m_{\mathrm{B}}$ receives quantum corrections leading to \cite{D03}
\begin{equation}
V_{Q}(r)=-{G m_{\mathrm {A}}m_{\mathrm {B}}\over r} \left[ \left(1+{k_{1}\over r}+{k_{2}\over r^{2}} \right) 
+{\rm O}(G^{2}) \right],
\label{1}
\end{equation}
where $k_{1} \equiv \kappa_{1} \left(L_{\mathrm{A}}+L_{\mathrm{B}} \right)$ and  $k_{2} \equiv \kappa_{2}{G \hbar \over c^{3}}=\kappa_{2}(l_{\mathrm{P}})^{2}$ ($L_{\mathrm{A}}$ and $L_{\mathrm{B}}$ being the gravitational radii of the bodies) \cite{BE14a,BE14b,BEDS15}. Throughout the paper we are going to use the numerical values \cite{D03} $\kappa_{1}=3$ and $\kappa_{2}={41 \over 10 \pi}$. Some of us\cite{BE14a} considered the application of Eq. (\ref{1}) to the circular restricted three-body problem of celestial mechanics, in which two bodies $\mathrm{A}$ and $\mathrm{B}$ of masses $\alpha$ and $\beta$, respectively, with $\alpha > \beta$, move in such a way that their orbits around their center of mass C are circular, while a third body, the planetoid P of mass $m$ much smaller than $\alpha$ and $\beta$, is subject to their gravitational attraction. On taking rotating axes $x-y$ with angular velocity $\omega$  and centred at C, one assumes that the quantum corrected effective potential for the circular restricted three-body problem is given by $GU$, with \cite{BE14a}
\begin{equation}
U={1\over 2}{(\alpha+\beta)\over l^{3}}(x^{2}+y^{2})
+{\alpha \over r} \left(1+{k_{1}\over r}+{k_{2}\over r^{2}}\right)
+{\beta \over s} \left(1+{k_{3}\over s}+{k_{2}\over s^{2}}\right),
\label{6}
\end{equation}
where $r$ and $s$ are the distances AP and BP, respectively, and $l$ indicates the distance AB. The equilibrium points of the system are found by studying the gradient of $U$ and evaluating its zeros. The simple but nontrivial idea in Refs.  \cite{BE14a,BE14b,BEDS15} was that, even though the quantum corrections in (\ref{1}) involve small quantities, the analysis of stable equilibrium (to linear order in perturbations) might lead to testable departures from Newtonian theory, being related to the gradient of $U$, and to the second derivatives of $U$ evaluated at the zeros of ${\rm grad}U$. By developing this idea, the works in Refs. \cite{BE14b,BEDS15}  have found, both for collinear Lagrangian points (i.e. $L_1$, $L_2$ and $L_3$) and noncollinear Lagrangian points (i.e., $L_4$ and $L_5$) in the Earth-Moon system, quantum corrections of the order of few millimetres, which represent, for the first time in the context of quantum gravity, effects that can be observed with the help of modern laser ranging techniques. 

\section{Quantum Effects on Lagrangian Points}

\subsection{Noncollinear Lagrangian points}

The ${\rm grad}U=0$ condition at noncollinear libration points leads to the algebraic equations of fifth degree \cite{BE14a}
\begin{equation}
w^{5}+\zeta_{4}w^{4}+\zeta_{3}w^{3}+\zeta_{0}=0, \;\; \; \; \;  \; \; \; \; \; u^{5}+{\tilde \zeta}_{4}u^{4}+{\tilde \zeta}_{3}u^{3}+{\tilde \zeta}_{0}=0,
\label{11}
\end{equation}
where we have set $w \equiv 1/r$ and $u \equiv 1/s$ and with
\begin{equation}
\zeta_{4}={2 \over 3}{\kappa_{1}\over \kappa_{2}}
{G(m+\alpha)\over c^{2}l_{P}^{2}}, \;
\zeta_{3}={1 \over 3 \kappa_{2}}{1 \over l_{P}^{2}}, \;
\zeta_{0}=-{1 \over 3 \kappa_{2}}{1 \over l_{P}^{2}l^{3}},
\label{13}
\end{equation}
\begin{equation}
{\tilde \zeta}_{4}={2 \over 3}{\kappa_{1}\over \kappa_{2}}{G(m+\beta)\over c^{2}l_{P}^{2}}, \;
{\tilde \zeta}_{3}=\zeta_{3}, \;
{\tilde \zeta}_{0}=\zeta_{0}.
\label{14}
\end{equation}
We have solved Eq. (\ref{11}) analytically by exploiting the full theory of the quintic equation, i.e., its reduction to
Bring-Jerrard form and the resulting expression of roots in terms of generalized hypergeometric functions \cite{BEDS15}.  In this way we have obtained the coordinates of quantum corrected noncollinear Lagrange points of the Earth-Moon system, finding that
\begin{equation}
x_{Q}-x_{C} \approx 8.7894 \; {\rm mm}, \;
|y_{Q}|-|y_{C}| \approx -4 \; {\rm mm},
\label{15}
\end{equation} 
Therefore, Eq. (\ref{15}) shows that to the equilateral libration points of Newtonian celestial mechanics there correspond points no longer exactly at vertices of an equilateral triangle.

\subsection{Collinear Lagrangian points}

For the collinear Lagrangian points, being characterized by the condition $y=0$, it is possible to relate distances $r$ and $s$ through the relation
\begin{equation}
s^{2}=(r -\varepsilon l)^{2} \Longrightarrow s= \pm (r -\varepsilon l), \; \; \; \varepsilon=\pm 1. 
\label{16}
\end{equation}
Depending on whether the sign of $s$ is positive or negative, equation (\ref{16}) leads to two algebraic ninth degree equations describing the position of $L_1$, $L_2$ and $L_3$. If we choose the positive sign in (\ref{16}), we obtain the equation \cite{BEDS15}
\begin{equation}
\sum_{n=0}^{9}A_{n}\psi^{n}=0,
\label{17}
\end{equation}
where the unknown is $\psi \equiv r/l$ and the coefficients $A_{n}$ read as
\begin{equation}
A_{0} \equiv =-3 (1+\rho)^{-1}\kappa_{2}(\rho_{P})^{2}, \; \; \; \; A_{1} \equiv -2(1+\rho)^{-1}\Bigr[\kappa_{1}\rho_{\alpha}-6 \varepsilon \kappa_{2}(\rho_{P})^{2}\Bigr],
\end{equation}
\begin{equation}
A_{2} \equiv -(1+\rho)^{-1}\Bigr[1-8 \varepsilon \kappa_{1}\rho_{\alpha}
+18 \kappa_{2} (\rho_{P})^{2}\Bigr],
\end{equation}
\begin{equation}
A_{3} \equiv 4(1+\rho)^{-1}\Bigr[\varepsilon -3 \kappa_{1}\rho_{\alpha}
+3 \varepsilon \kappa_{2}(\rho_{P})^{2}\Bigr],
\end{equation}
\begin{equation}
A_{4} \equiv -(1+\rho)^{-1}\Bigr[(6+(1+\varepsilon)\rho)
-2 \kappa_{1}(4 \rho_{\alpha}+\rho_{\beta}\rho)\varepsilon
+3(1+\rho)\kappa_{2}(\rho_{P})^{2}\Bigr],
\end{equation}
\begin{equation}
A_{5} \equiv (1+\rho)^{-1}\Bigr[(1+4 \varepsilon) +(5+ 2 \varepsilon)\rho
-2 \kappa_{1}(\rho_{\alpha}+\rho_{\beta}\rho)\Bigr],
\end{equation}
\begin{equation}
A_{6} \equiv -(1+\rho)^{-1}\Bigr[(1+4 \varepsilon)+(10 \varepsilon+1)\rho\Bigr], \; \; \; \; A_{7} \equiv 2(1+\rho)^{-1}(3+5 \rho),
\end{equation}
\begin{equation}
A_{8} \equiv -(1+\rho)^{-1}(4+5 \rho)\varepsilon, \; \; \; \; A_{9} \equiv 1, 
\end{equation}
where we have set
\begin{equation}
\rho \equiv {\beta \over \alpha}, \; 
\rho_{\alpha} \equiv {l_{\alpha}\over l}, \; \rho_{\beta} \equiv {l_{\beta}\over l}, 
\; \rho_{P} \equiv {l_{P}\over l},
\end{equation}
$l_{\alpha}$ and $l_{\beta}$ being the sum of the gravitational radii of the Earth and the planetoid and of the Moon and the planetoid, respectively. On the other hand, if we choose $s=-(r- \varepsilon l)$ in (\ref{16}), we end up with the algebraic equation \cite{BEDS15}
\begin{equation}
\sum_{n=0}^{9}B_{n}\psi^{n}=0,
\label{26}
\end{equation}
where
\begin{equation}
B_{k}=A_{k} \; {\rm if} \; k=0,1,2,3,7,8,9,
\end{equation}
\begin{equation}
B_{4} \equiv (1+\rho)^{-1}\Bigr[(-6+(1-\varepsilon)\rho)
+2 \kappa_{1}\varepsilon (4 \rho_{\alpha}+\rho_{\beta}\rho)
+3(\rho-1)\kappa_{2}(\rho_{P})^{2}\Bigr],
\end{equation}
\begin{equation}
B_{5} \equiv (1+\rho)^{-1}\Bigr[(1+4 \varepsilon)+(5 -2 \varepsilon)\rho 
-2 \kappa_{1}(\rho_{\alpha}+\rho_{\beta}\rho)\Bigr],
\end{equation}
\begin{equation}
B_{6} \equiv -(1+\rho)^{-1}\Bigr[(1+4 \varepsilon)+(10 \varepsilon -1)\rho].
\end{equation} 
In the classical limit $k_1=k_2=k_3=0$, Eqs. (\ref{17}) and (\ref{26}) get reduced to a pair of fifth degree algebraic equations (see Ref. \cite{BEDS15} for furthed details). By virtue of (\ref{17}), (\ref{26}) and of the two classical fifth degree equations mentioned above, we have found, for $L_1$, $L_2$ and $L_3$, respectively, that
\begin{equation}
R_{1}-r_{1}=3.6 \; {\rm mm}, \; \; R_{2}-r_{2}=2.3 \; {\rm mm}, \; \; R_{3}-r_{3}=9 \; {\rm mm},  \label{31}
\end{equation}
where $R_i$ represent the roots of the nonic equations (\ref{17}) and (\ref{26}), whereas $r_i$ are the corresponding classical values. Interestingly, the order of magnitude of quantum corrections to the location of $L_{1},L_{2},L_{3}$ in the Earth-Moon system coincides with the order of magnitude of quantum corrections to $L_{4},L_{5}$. This may not have any practical consequence, since $L_{1},L_{2},L_{3}$ are points of unstable equilibrium, but the analysis performed in this
section adds evidence in favour of our evaluation of quantum corrections to all Lagrangian points in the Earth-Moon system being able to predict effects of order half a centimeter.

\section{The Laser Ranging Technique}

The quantum gravity effects described so far can be studied with the technique of Satellite/Lunar Laser Ranging and a laser-ranged test mass equipped with 
Cube Corner Retro-reflectors, to be designed ad hoc for this purpose. Some of the Key Performance Indicators that must be taken into account to design an appropriate laser-ranged test mass for the signature of new physics described here are as follows.
\begin{itemize}
\item[(i)] Adequate laser return signal from the Lagrangian  points $L_{4},L_{5}$.
\smallskip
\item[(ii)] Acceptable rejection of the unavoidable nongravitational perturbations at $L_{4},L_{5}$ which any chosen test mass and/or 
test spacecraft will experience.
\smallskip
\item[(iii)] Optimization/minimization of the value of the surface-to-mass ratio, S/M. This is a critical point, since all nongravitational perturbations related to the sun radiation pressure and thermal effect, are proportional to S/M.
\smallskip
\item[(iv)] Time-durability of the test mass to prolonged measurements.
\end{itemize}

\section{Conclusions}

Our analysis shows that the quantum corrections on the position of the planetoid at all Lagrangian points are of order of a few millimetres, as witnessed by Eqs. (\ref{15}) and (\ref{31}). Therefore our model, unlike all other theories of quantum gravity, produces predictions which are testable with the help of laser ranging technique. This is an astonishing result because it involves a really familiar and approachable system like the Earth-Moon system. Finally, we have to consider that a modern version of the planetoid is the solar sail, and much progress has been made, in recent years, on the displaced periodic orbits of solar sails at all libration points, both stable and unstable. The work in Ref. \cite{BEDS15} has shown that displaced periodic orbits for the solar sail exist in the Earth-Moon system also when the quantum potential (\ref{6}) is assumed. By solving the variational equations of motion at the libration points $L_4$ and $L_5$, we have found that, both in the Newtonian and in the quantum case, trajectories around Lagrangian points are ellipses centred at the Lagrangian points themselves, showing that also in the quantum regime periodic solutions in the neighborhood  of uniform circular motion are possible \cite{BEDS15}.

\end{document}